\input harvmac
\hsize=16.5truecm 
\vsize=23.5truecm 
\overfullrule=0pt
\parskip=3pt
 at 14truept
\overfullrule 0pt
\voffset -10pt 


\def\hexnumber@#1{\ifcase#1 0\or1\or2\or3\or4\or5\or6\or7\or8\or9\or
        A\or B\or C\or D\or E\or F\fi }

%
\def\boxit#1{\leavevmode\kern5pt\hbox{
	\vrule width.2pt\vtop{\vbox{\hrule height.2pt\kern5pt
        \hbox{\kern5pt{#1}\kern5pt}}
      \kern5pt\hrule height.2pt}\vrule width.2pt}\kern5pt}

\def\cH{{\cal H}}
\def\cN{{\cal N}}
\def\cM{{\cal M}}
\def\mod{{\rm mod}\ }
\def\id{{\rm id}}
\def\s{{\overline s}}
\def\lv{\left|\,}
\def\rv{\,\right|}
\def\lb{\left(\,}
\def\rb{\,\right)}

\def\\{\hfil\break}
\def\la{\lambda}

\def\Z{{\bf Z}}

\def\b#1{\big(#1\big)}

\def\+{\oplus}
\def\x{\otimes}
\def\r{{\overline r}}

\def\b#1{\kern-0.25pt\vbox{\hrule height 0.2pt\hbox{\vrule
width 0.2pt \kern2pt\vbox{\kern2pt \hbox{#1}\kern2pt}\kern2pt\vrule
width 0.2pt}\hrule height 0.2pt}}

\def\STrow#1{\hbox{#1}\kern-1.35pt}

\font\huge=cmr10 scaled \magstep2
\font\small=cmr8


{\nopagenumbers
\rightline{DAMTP-98-18}
\rightline{hep-th/9804040}
\vskip .75cm
\centerline{{\huge\bf Heterotic Modular Invariants and Level--Rank Duality}
{\footnote*{\small Supported in part by NSERC.}}}
\bigskip
\bigskip\centerline{T. Gannon\ \footnote{$^\ddagger$}{\small Address
from July 1998: Math Dept, Univ.\ Alberta, Edmonton, Canada} and M.A.
Walton\
\footnote{$^\dagger$}{\small On leave from the Physics Dept, Univ. Lethbridge,
Alberta, Canada}}

\bigskip
\centerline{{\it Department of Mathematics,
York University}} \centerline{{\it Toronto, Ontario,
Canada\ \ M3J 1P3}}
\centerline{tgannon@mathstat.yorku.ca}

\bigskip
\centerline{\it Department of Applied Mathematics and Theoretical Physics}
\centerline{\it 
University of Cambridge} \centerline{{\it Silver Street,
Cambridge\ \ CB3 9EW, U.K.}}\centerline{\rm m.a.walton@damtp.cam.ac.uk} 

\vskip .75cm \centerline{{\bf Abstract}} \noindent{New heterotic
modular invariants are found using the level-rank duality of affine
Kac-Moody algebras. They provide strong evidence for the consistency
of an infinite list of heterotic Wess-Zumino-Witten (WZW) conformal
field theories. We call the basic construction the {\it dual-flip},
since it flips chirality (exchanges left and right movers) and takes
the level-rank dual. We compare the dual-flip to the method of
conformal subalgebras, another way of constructing heterotic
invariants. To do so, new level-one heterotic invariants are first
found; the complete list of a specified subclass of these is
obtained. We also prove (under a mild hypothesis) an old conjecture
concerning exceptional $A_{r,k}$ invariants and level-rank duality.}

\leftskip=0cm \rightskip=0cm
\vskip.5cm
\noindent PACS:\ \ 11.25.Hf,\ 02.20.Tw\hfill\break
\vfill

\eject}

\pageno=1
\newsec{Introduction}

Many methods are now known for the direct construction of 
Wess-Zumino-Witten (WZW) modular invariants. The most important among these
start from generic ``symmetries'' of the modular matrix $S$.

For example, the WZW models realise affine Kac-Moody algebras as their
current algebras. Symmetries of $S$ result from diagram
automorphisms of the affine algebras that are not also diagram
automorphisms of their horizontal (finite-dimensional) subalgebras.
Orbifolds by these
symmetries have  non-diagonal modular invariants as their torus
partition functions \ref\gepw{D. Gepner and E. Witten, {\sl
Nucl. Phys.} {\bf B278} (1986) 493}. (In the more general context of
rational conformal field theory, this orbifold procedure is known as
the method of simple currents \ref\sy{A.N. Schellekens and
S. Yankielowicz, {\sl Nucl. Phys.} {\bf B327} (1989)
673}\ref\int{K. Intriligator, {\sl Nucl. Phys.} {\bf B332} (1990)
541}.)

The WZW $S$ matrices also have nice arithmetic properties, and so obey
certain Galois relations \ref\cg{A. Coste and T. Gannon, {\sl
Phys. Lett.} {\bf B323} (1994) 316}. In \ref\fss{J. Fuchs, B.
Gato-Rivera, B. Schellekens and C. Schweigert, {\sl Phys. Lett.} {\bf
B334} (1994) 113}, these Galois relations were used to construct
modular invariants. As for simple currents, one can generalise the
Galois considerations to arbitrary rational conformal field theories.

One generic symmetry of WZW modular $S$ matrices sticks out,
however. The level-rank duality of WZW models \ref\kn{A. Kuniba and
T. Nakanishi, in:\hfill\break S. Das et al. (editors), {\sl Modern
Quantum Field Theory} (World Scientific, New York, 1991) (proceedings
of the International Colloquium on Modern Quantum Field Theory,
Bombay, 1990)}\ref\mnrs{E. Mlawer, S. Naculich, H. Riggs and
H. Schnitzer, {\sl Nucl. Phys.}  {\bf B352} (1991)
863}\ref\nt{T. Nakanishi and A. Tsuchiya, {\sl Comm. Math. Phys.} {\bf
144} (1992) 351} has not been formulated in a general rational
conformal field theory context. More importantly here, it has not been
used in a direct manner to construct WZW modular
invariants\foot{Level-rank duality has previously been used in an
indirect manner to construct modular invariants \ref\mw{M.A. Walton,
{\sl Nucl. Phys.} {\bf B322} (1989) 775}\ref\dv{D. Verstegen, {\sl
Comm. Math. Phys.} {\bf 137} (1991) 567}.}. We correct this latter
omission.

What we find are new {\it heterotic} modular invariants, i.e. those
describing theories with different holomorphic (left-moving) and
anti-holomorphic (right-moving) excitations (see \ref\tghet{T. Gannon,
{\sl Nucl. Phys.} {\bf B402} (1993) 729}\ref\tgqhk{T. Gannon and
Q. Ho-Kim, {\sl Nucl. Phys.} {\bf B425} (1994) 319}).  The modular
invariants are thus integer (sesquilinear) combinations of characters
of one affine algebra with complex conjugates of the characters of a
{\it different} (possibly trivial) algebra. For example, in simple
cases, the invariants will be linear combinations of the characters of
affine algebras, describing systems -- the so-called {\it meromorphic}
conformal field theories -- with no right-moving part.  In
these cases, the construction can be indicated schematically as
\eqn\spec{(g;g)\ \ \Rightarrow\ \ \ (g\oplus\tilde g;\{\})\, \ \ .}
Here $g$ indicates an affine nontwisted Kac-Moody algebra $X_{r,k}$,
where $X_r$ is the simple Lie horizontal subalgebra of rank $r$, and
$k$ is the fixed level. The semi-colon separates the algebras of the
left-moving ($L$) and right-moving ($R$) sectors, and $\tilde g$ is
the algebra that is level-rank dual to $g$.

To see why, notice that level-rank duality equates the $S$ and $T$
matrices of one WZW model to the {\it complex conjugates} of the $S$
and $T$ matrices of the dual WZW model (see below), rather than to the
dual $S$ and $T$ matrices themselves. To remedy this, we need a way of
exchanging the complex conjugates for the dual modular matrices. But
this kind of thing has been done in \ref\ganw{T. Gannon and
M.A. Walton, {\sl Comm. Math. Phys.} {\bf 173} (1995) 175}, where the
modular invariants for certain diagonal coset theories were
classified. There a `switch' (or `flip') of left-moving and
right-moving fields was used in order to relate the modular
transformations of fields of the coset theory to those of the
corresponding direct product theory. One of the ways elements of the
coset and direct product modular matrices differ is by complex
conjugation applied to one factor matrix.

Here we use a switch similar to that of \ganw\ on theories with an
affine left-moving current algebra and the same algebra on the
right-moving side. Consequently, heterotic invariants are obtained
with current algebra plus dual current algebra that are both
left-moving (say). In the simplest case then, a purely holomorphic
invariant is obtained, as in \spec. The most general form of the
construction is \eqn\gen{(h_L; h_R)\ =\ (g_1\oplus g_2;g_3\oplus g_4)\
\
\Rightarrow\ \ \ (g_L;g_R)\ =\ (g_1\oplus 
\tilde g_4;{g}_3\oplus \tilde{g}_2)\,
\ \ .} Here $h_{L(R)}$ and $g_{L(R)}$ indicate the original and final
left(right)-moving algebras, respectively. The symbols $g_i$ can now
stand for direct sums of affine nontwisted Kac-Moody algebras.

The key result is that a combination of a level-rank duality
transformation and a chirality flip (changing right-moving to
left-moving) can map a modular invariant to another, new modular
invariant. For short, we will refer to this operation as a {\it
dual-flip}. The lesson seems to be that a level-rank duality
transformation is naturally regarded as also exchanging left-movers
and right-movers (`level$\leftrightarrow$rank' means
`Left$\leftrightarrow$Right').

The construction does not work as generally as one might think by
reading this far, however. Nevertheless, it seems to be a major source
of heterotic WZW modular invariants. Since the heterotic modular
invariants are generally more difficult to construct than
nonheterotic (left-right symmetric) ones \tghet\tgqhk, this is
significant progress.

We find meromorphic (i.e.\  $(g_L;g_R)=(g_L;\{\})$) invariants
for\smallskip

\item{$\bullet$} $g_L=su(n)_m\oplus su(m)_n$ whenever $mn$ is a perfect
square congruent to 1 (mod 24);

\item{$\bullet$} $g_L=C_{r,k}\oplus C_{k,r}$ whenever 12 divides $rk$;

\item{$\bullet$} $g_L=B_{r,k}\oplus D_{{k\over 2},2r+1}$ whenever 48 divides
$(2r+1)k$;

\item{$\bullet$} $g_L=D_{r,k}\oplus D_{{k\over 2},2r}$ whenever $k$ is even
and 24 divides $rk$.\smallskip 

The method of conformal subalgebras \ref\bt{A. Bais and A. Taormina,
{\sl Phys. Lett.} {\bf B181} (1986) 87} provides the only other known
general source of heterotic modular invariants.  By comparing it with
our {\it dual-flip method} we find more new results.  Specifically, we
show that some of our invariants can also be derived by conformal
embeddings  followed by modular invariant contraction \ref\bou{P. Bouwknegt,
{\sl Nucl. Phys.} {\bf B290} (1987) 507}. But this is only possible
once we have constructed previously unknown {\it level-one} heterotic
physical invariants (recall that only level-one nontwisted affine
algebras can have conformal subalgebras). 

In the simplest
case, these level-one invariants are meromorphic, and so cannot
be constructed by the dual-flip. We classify
all invariants corresponding to algebras $(X_{r,1};\{\})$,
$(X_{r,1};X_{s,1})$ and $(X_{r,1}\oplus X_{s,1};\{\})$. Invariants
obtained from these by conformal embeddings, followed by contraction 
with other invariants, turn out to be of the type derived from
the dual-flip. It seems probable that {\it all} dual-flip invariants can
be derived in this fashion, once the appropriate level-one invariant
is constructed.

Among the specific examples we find are $c=24$ meromorphic
invariants. These can be related, by conformal embeddings and
conjugations, to entries on Schellekens' list
\ref\schell{A.N. Schellekens, {\it Comm. Math. Phys.} {\bf 153} (1993)
159}. 

A major result of this paper is our proof of the conjecture of \mw\
concerning the relation between nonheterotic exceptionals of $h_L=h_R=
su(\r)_k$ and $g_L=g_R=su(k)_{\r}$. There is no bijection between the
sets of such physical invariants (unlike the simpler situation for
$C_{r,k}\leftrightarrow C_{k,r}$), but under a mild technical
condition which we expect to always be satisfied, $su(\r)_k$ will have
exceptionals iff $su(k)_{\r}$ will. This is discussed toward the end of
section 5.

We expect that most, if not all, of the physical invariants we find
are partition functions of sensible WZW conformal field theories,
whether or not they are heterotic. In order to provide further
evidence that such theories are consistent, however, one should
calculate four-point functions, for example, and verify they satisfy
the required properties. Of course, some of the modular invariants we
find are the partition functions of known conformal field theories 
\ref\dgm{ l. Dolan, P. Goddard and P. Montague, {\sl Phys. Lett.}
{\bf B236} (1990) 165}.

What we do here, though, is construct new invariants from known ones,
many of which are undoubtedly the partition functions of consistent
theories. And we believe that since level-rank duality is a
duality between the spaces of conformal blocks of theories
\ref\cblocks{S. Naculich and H. Schnitzer, {\sl Nucl. Phys.} {\bf
B347} (1990) 687;\hfill\break T. Nakanishi and A. Tsuchiya, {\sl
Comm. Math, Phys.} {\bf 144} (1992) 351}, the dual-flip extends from a
map between invariants to a map between theories.

Section 2 treats preliminaries and the construction for the $A_r$ WZW
models. In section 3 the results are extended to the other classical
simple Lie algebras.  Section 4 contains some explicit examples. The
full list of heterotic modular invariants for algebras
$(X_{r,1};\{\})$, $(X_{r,1};X_{s,1})$ and $(X_{r,1}\oplus
X_{s,1};\{\})$ (for all simple $X$) is given in section 5, where
it is also verified (in certain cases only) that invariants generated
from them are of the `dual-flip type'. A conclusion is given in
section 6.

\newsec{Preliminaries and $A_r$ invariants}

Let $X_r^{(1)}$ denote the nontwisted affine Kac-Moody algebra that is
the central extension of the loop algebra of the simple Lie algebra
$X_r$.  As indicated above, we use $X_{r,k}$ to signify this  affine
algebra at fixed positive-integer level $k$. The corresponding set of
integrable highest weights has the following set of horizontal
projections: \eqn\inthw{P_+(X_{r,k})=\{\ \la=\sum_{i=1}^r \la_i w^i\ \
:\ \ \la_i\in\Z_{\ge0},\ \sum_{i=1}^r \la_i a^\vee_i \le k\ \}\ \ .}
Here the $a^\vee_i$ are co-marks, and $w^i$ denotes the $i$-th
fundamental weight of $X_r$. It will often be convenient to introduce the
`0th affine Dynkin label' $\la_0$ defined by \eqn\Dlzero{\la_0:=
k-\sum_{i=1}^r\la_ia_i^\vee\ .}

The affine characters ${\rm ch}_\la(\tau,z,u)$ transform covariantly
under the action of the modular group ${\rm SL}_2(\Z)$. We have
\eqn\Ttransf{{\rm ch}_\la(\tau+1,z,u)\ =\ \sum_{\mu\in P_+(X_{r,k})}\
T_{\la}^{\mu\,}\ {\rm ch}_\mu(\tau,z,u)\ =\ \ T_{\la}^{\la}\ {\rm
ch}_\la(\tau,z,u)\ \ ,} and \eqn\Stransf{{\rm
ch}_\la(-1/\tau,z/\tau,u-z\cdot z/2\tau)\ =\ \sum_{\mu\in P_+(X_{r,k})}
S_{\la}^{\mu}\ {\rm ch}_\mu(\tau,z,u)\ \ .}

The modular matrices $S$ and $T$ are symmetric and unitary, and have
many other interesting properties (see \ref\gwfr{T. Gannon and
M.A. Walton, {\it On fusion algebras and modular matrices},
q-alg/9709039 (1997)}, e.g.). For our purposes, the most important
relate to the symmetries of the (extended) Coxeter-Dynkin diagram of
$X_{r}^{(1)}$, and to level-rank duality. The diagram symmetries act
on weights by permuting their Dynkin labels. There is the involutive
charge conjugation $C$, which is a symmetry of the (unextended)
Coxeter-Dynkin diagram of $X_r$, and so fixes the zeroth node of the
$X_{r}^{(1)}$ diagram. The other important diagram symmetries are
often called simple currents, and are related to the centre of the Lie
group $\exp(X_r)$ one gets by exponentiating $X_r$. When the simple
current group is generated by a single element, we will denote the
generator by $J$, and any element of that group by $A$. The only
exception is $X_r=D_r$, with $r$ even. For that case, the two
generators will be represented by $J$ and $J_s$, with the subscript
indicating spinor. One finds (see
e.g. \ref\ber{D. Bernard, {\sl Nucl. Phys.}  {\bf B288} (1987) 628})
\eqn\JSt{S_{\la}^{A\mu}\ =\ S_{\la}^{\mu}\ e^{2\pi
i(A^{-1}w^0)\cdot\la}\ =\ S_{\la}^{\mu}\ e^{2\pi it_A(\la)/|A|}\ \ ,}
where $|A|$ is the order of $A$, and
$t_A(\la):=|A|\,(A^{-1}w^0)\cdot\la$ is an integer obtained from the
Dynkin labels of the weight $\la$. The phase in \JSt\ is the
eigenvalue of an element of the centre of $\exp(X_r)$, with any vector
in the representation of highest weight $\la$ as eigenvector.

For example, with $X_{r,k}=A_{r,k}$, we have the diagram symmetry $J$
that permutes the fundamental weights in a cyclic manner, so that
\eqn\oaA{J\la\ =\ J(\sum_{i=1}^r\, \la_i w^i)\ =\ \sum_{i=1}^r
\la_{i-1}w^i\ . } 
$J$ has order $|J|=r+1=:\r$, and $t_J(\la)$ in
\JSt\ is the $\r$-ality \eqn\tA{t_J(\la)\ =\ \r(J^{-1}w^0)\cdot\la\ =\
\r w^r\cdot\la\ =\ \sum_{i=1}^r\, i\la_i\ \ .}  It behaves nicely
under the action of $J$: \eqn\tJA{t_J(J^a\la)\ \equiv\ ka+t_J(\la)\
({\rm mod}\ \r)\ \ .}  Using this, and \eqn\tJaJ{t_{J^a}(\la)/|J^a|\
\equiv\ t_J(\la)/|J|\ ({\rm mod}\ 1)\ ,} one can derive a
generalisation of \JSt: \eqn\JStA{S_{J^a\la}^{J^b\mu}\ =\ \exp[2\pi
i(bt_J(\la)+at_J(\mu)+kab)/\r]\ S_{\la}^{\mu}\ \ .} The action of $J$
on the elements of $T$ is given by \eqn\JTA{T_{J^a\la}^{J^a\mu}\ =\
\exp[\pi i(-2at_J(\la)+ka(\r-a))/\r]\ T_{\la}^{\mu}\ \ .}

Many (and sometimes all) highest weights of $X_{r,k}$, with $X_r$
classical, can be described by Young diagrams. The transpose of a
Young diagram corresponds to a highest weight of a different algebra,
where the roles of rank and level are interchanged. If $\la$ is the
original highest weight, we denote the highest weight corresponding to
the transposed diagram by $R\la$. The relations between elements of
the modular matrices are not so easily described in a uniform, general
way. Restricting to $X_{r,k}=A_{r,k}$ then, we find
\ref\abi{D. Altschuler, M. Bauer and C. Itzykson, {\sl
Comm. Math. Phys.} {\bf 132} (1990) 349} that the map $R$ is not a
bijection between the sets of weights $P_+(A_{r,k})$ and
$P_+(A_{k-1,r+1})$. But it is a bijection between the
$J$-orbits of $P_+(A_{r,k})$ and the $\tilde J$-orbits of
$P_+(A_{k-1,r+1})$. (Tildes denote objects relevant to the algebra
$A_{k-1,r+1}=su(k)_\r$ that is level-rank dual to $A_{r,k}=su(\r)_k$.)
More importantly here, one can also construct a bijection $R_0$
between a subset of $P_+(A_{r,k})$ and a subset of
$P_+(A_{k-1,r+1})$. We define the subsets of weights
\eqn\Pzero{P^0_+(A_{r,k}):=\{\ \la\in P_+(A_{r,k})\ \ :\ t_J(\la)\equiv 0\
(\mod \r) \}\ \ .} It can be verified that $R_0$ given by
\eqn\Rzero{R_0\,\lambda\ =\ \tilde J^{-t_J(\la)/\r}\, R\la\ } is a bijection
between $P^{0}_+(A_{r,k})$ and $\tilde
P^{0}_+(A_{k-1,r+1})$. Furthermore
\ref\tgades{T. Gannon, {\it Kac-Peterson, Perron-Frobenius, and the
classification of conformal field theories}, q-alg/9510026
(1995)}, \eqn\lrSTAprime{S_{\la}^{\mu}\ =\ \sqrt{{k\over \r}}\, \tilde
S_{R_0\la}^{*\,R_0\mu}\ \ \ ,\ \ \ \ T_{\la}^{\mu}\, T_{0}^{*\,0}\ =\ \tilde
T_{R_0\la}^{*\,R_0\mu}\, \tilde T_{\tilde 0}^{\tilde 0}\ \ , } for all
$\la,\mu\in P^0_+(A_{r,k})$.

In general, the partition function of a WZW model will take the form
of a sesquilinear combination of affine characters: \eqn\genz{{\cal
H}(\tau,z,u)\ =\ \sum_{\mu\in P_+}\ \sum_{\nu'\in P_+'}\ {\rm ch}_\mu
\, H_\mu^{\nu'}\, {\rm ch}^*_{\nu'}\ \ .} Notice that we use lower
indices for `left-movers' (holomorphic sector) and upper indices for
`right-movers' (anti-holomorphic sector). The primes emphasise that
the affine algebras may be different on the two `sides' of the
theory. For example, $P_+$ and $P_+'$ denote the sets of integrable
highest weights of the two algebras. If the two affine algebras
coincide, the partition function describes a nonheterotic theory. If
they do not, it is relevant to a heterotic theory, as in \genz. If
only the left-movers (right-movers) are present, we have a meromorphic
(anti-meromorphic) theory. We will use the symbols $\cH,H$; $\cN,N$;
and $\cM,M$ to indicate heterotic; nonheterotic; and meromorphic
modular invariants, respectively.

By a {\it physical invariant} we mean a combination of characters of the form
\genz, with $H_{\mu}^{\nu'}\in \Z_{\ge 0}$ and $H_0^{0'}=1$,
satisfying\foot{This term is now standard, though somewhat misleading:
note that a {\it physical} invariant might not be realised as the partition
function of a consistent theory.}
\eqn\Sinv{\sum_{\la}\, S_{\mu}^{\la}\, H_{\la}^{\nu'}\ =\ \sum_{\la'}\,
H_{\mu}^{\la'}\, S_{\la'}^{\prime\,\nu'}\ \ ,} and
\eqn\Tinv{\sum_{\la}\, T_{\mu}^{\la}\, H_{\la}^{\nu'}\ =\
\sum_{\la'}\, H_{\mu}^{\la'}\, T_{\la'}^{\prime\,\nu'}\ \ .}  In
\Tinv, the condition $H_0^{0'}=1$ requires \eqn\clcr{c\ \equiv\ c'\
({\rm mod}\ 24)\ \ ,} which ensures there is no modular anomaly. Here
$c,c'$ are the central charges of the two sides of the theory. For
example, if the affine algebra on one side is $X_{r,k}$ with $X_r$
simple, then the corresponding central charge is
$c(X_{r,k})=kD/(k+h^\vee)$, where $D$ is the dimension of $X_r$, and
$h^\vee$ is its dual Coxeter number.

It is of some value to study candidate partition functions that
satisfy \Sinv\ exactly, but satisfy \Tinv\ only when $T_{\mu}^{\la}$
and $T_{\la'}^{\prime\,\nu'}$ there are replaced with
$T_{\mu}^{\la}\,T_{0}^{*\ 0}$ and
$T_{\la'}^{\prime\,\nu'}\,T_{0'}^{\prime\,*\,0'}$ respectively (so the
modular anomaly condition \clcr\ may not be satisfied). We call these
objects {\it anomalous} physical invariants. They can appear as the
partition functions of certain sectors of critical string partition
functions. For example, the sector of {\it one} of the factors $E_8$
in the $E_8\x E_8$ heterotic string is described by such a partition
function\foot{In heterotic string phenomenology, one $E_8$ part of the
theory can play the role of a {\it hidden sector}
\ref\chsw{P. Candelas, G. Horowitz, A. Strominger, and E. Witten, {\sl
Nucl. Phys.} {\bf B258} (1985) 46}.  The possible relevance of
heterotic (anomalous) physical invariants to hidden sectors was pointed
out by C.S. Lam (as cited in \tghet).}. They will also be useful here
as simple examples to illustrate our construction.

The following simple observations will be used repeatedly in what
follows (proofs can be found in \tgades). Let $\cH$ be any physical
invariant (anomalous or otherwise). Let 
${\cal J}_{LR}$ be the set of all simple currents $(A;A')$, where $A$
is a simple current of the holomorphic sector, and $A'$ is a simple
current of the anti-holomorphic sector, and let ${\cal J}(H)$ be the
set of all $(A;A') \in{\cal J}_{LR}$ such that $H_{A0}^{A'0'}\ne 0$. Then
${\cal J}(H)$ is a subgroup of ${\cal J}_{LR}$, and for any
$(A;A')\in{\cal J}_{LR}$, \eqn\MJii{(A;A')\in{\cal J}(H)\
{\rm iff}\quad H_{A\la}^{A'\mu'}=H_{\la}^{\mu'}\ {\rm for\ all}\ \la\in
P_+,\ \mu'\in P_+'\ ,} and finally, $(A;A')\in{\cal J}(H)$ iff the
following selection rule holds: \eqn\MJiii{H_{\la}^{\mu'}\ne 0\quad{\rm
implies}\quad t_A(\la)/|A|\equiv t_{A'}(\mu')/|A'|\ ({\rm mod}\ 1)} for all
$\la\in P_+,\mu'\in P_+'$.

Our objective here is to write heterotic invariants
using level-rank duality.  In the simplest case, we construct a
meromorphic invariant from a known nonheterotic invariant. Let
$N_{\mu}^{\nu}$ and $M_{\mu,\tilde\nu}$ denote the coefficient
matrices of the original nonheterotic and resulting meromorphic invariant,
respectively. In the remainder of this section we will consider only
the $A_{r}$ algebras. Consider first the case where the affine algebra
of the invariant $\cN$ is $A_{r,k}$ for both left- and right-movers, and
the meromorphic invariant $\cM$ has algebra $A_{r,k}\oplus
A_{k-1,r+1}$. The simplest construction of a meromorphic invariant (we
will generalise this later in this section) corresponds to the choice
\eqn\HM{M_{\mu,\tilde\nu}\ =\ \left\{\matrix{ N_{\mu}^{R_0\tilde\nu}\
,&\ \ {\rm for\ all}\ \mu\in P_+^0(A_{r,k}),\ \tilde\nu\in \tilde
P_+^0(A_{k-1,r+1})\ ,\cr 0,&\ {\rm otherwise}\ \ .\cr}\right. } Notice
the similarity with the `switch' used in \ganw; there left- and
right-moving fields were exchanged, while here a right-moving field is
traded for a left-mover that is also its level-rank dual. Using the
term introduced above, fields are replaced by their flipped-duals.

The construction of \HM\ only works for those physical invariants
$\cN$ satisfying \eqn\Mcond{N_{\la}^{\mu}\not=0\ \ \Rightarrow\ \
\la,\mu\in P_+^0(A_{r,k})\ \ .}  But such invariants exist, and they
can also be characterised as obeying $N_{J0}^{0}=N_{0}^{J0}=1$ \MJiii.
We can find a useful general constraint right away then, by applying
$T$-invariance \Tinv, \JTA\ to $N_{J0}^{0}=1$: if $r$ is even, $\r$
must divide $k$; and if $r$ is odd, $2\r$ must divide $k$.

To verify modular invariance \Sinv, we must show that \eqn\SSH{\sum_{\la\in
P_+(A_{r,k})}\ \ \sum_{\tilde\zeta\in P_+(A_{k-1,r+1})}\,
S_{\mu}^{\la}\ \tilde S_{\tilde\nu}^{\tilde\zeta}\ M_{\la,\tilde\zeta}\
=\ M_{\mu,\tilde\nu}\ \ .} With our ansatz, the left hand side becomes
\eqn\SSMone{ \sum_{\la\in P^0_+(A_{r,k})}\ \ \sum_{\tilde\zeta\in
\tilde P^0_+(A_{k-1,r+1})}\, S_{\mu}^{\la}\ N_{\la}^{R_0\tilde\zeta}\ \tilde
S_{\tilde\nu}^{\tilde\zeta}\ ,} where we use \Mcond. Since $R_0$ is a
bijection of $P^0_+(A_{r,k})$ onto $\tilde P^0_+(A_{k-1,r+1})$, we can write
\eqn\SSMtwo{\sum_{\la,\zeta\in P^0_+(A_{r,k})}\, S_{\mu}^{\la}\ 
N_{\la}^{\zeta}\ \tilde S_{R_0\zeta}^{\tilde\nu}\ =\
\sum_{\la,\zeta\in P^0_+(A_{r,k})}\, S_{\mu}^{\la}\ N_{\la}^{\zeta}\,
\left( \sqrt{{\r\over k}}\, S^*_{\zeta,R_0\tilde\nu}\right)\ ,} using
the symmetry of $S$ and \lrSTAprime. By the modular invariance of
$N_{\mu}^{\nu}$, this is just $N_{\mu}^{R_0\tilde\nu} \sqrt{{\r\over
k}}$. The $S$-invariance of $M_{\mu,\tilde\nu}$ follows, if we have
\eqn\krbar{k\ =\ \r\ \ .}

$T$-invariance is shown in similar fashion, using \lrSTAprime. The
factors of $T_{0}^{*\,0}$ and $\tilde T_{\tilde 0}^{\tilde 0}$ are the only
complication. But they are absent if the central charge condition
\clcr\ is satisfied; here it takes the form \eqn\chetA{c(A_{r,k}) +
c(A_{k-1,r+1})\ =\ k(r+1)-1\ \equiv\ 0\,({\rm mod}\ 24)\ \ .}  The
condition therefore reduces to \eqn\cheta{(r+1)^2\ =\ k^2\ \equiv\ 1\
({\rm mod}\ 24)\ \ ,} after imposing \krbar. So the ansatz \HM\
produces a non-anomalous physical invariant iff $k=\r$ is odd and not
divisible by 3; \HM\ is anomalous iff $k=\r$ is odd and divisible by
3.

There is a way to weaken the stringent requirement \krbar, by generalising
\HM\ somewhat. $\r$ and $k$ must remain odd, but condition \krbar\ can be
relaxed to \eqn\perfsq{\r k\ {\rm is\ a\ perfect\ square.\ }} To see how,
write out the prime decomposition of $\r/k$ as $\prod_i
p_i^{2a_i}\prod_jq_j^{-2b_j}$, where $p_i,q_j$ are distinct primes,
and $a_i,b_j$ are positive integers. Define $a:=\prod_i p_i^{a_i}$,
$b:=\prod_j q_j^{b_j}$; then we have $\r/k=(a/b)^2$, with $a$ and $b$
two coprime integers.

The key is to replace \Mcond\ with  \eqn\Mcondw{N_{\la}^{\mu}\ne 0\
\Rightarrow\ t_J(\la)\equiv t_J(\mu)\equiv 0\ (\mod \r/a).\ }
Equivalently, we need only demand $N_{J^a0}^{0}=N_{0}^{J^a0}=1$, by \MJiii.
Such $\cN$ will always exist, for
the given choice of $a$ (provided $\r$ is odd and the perfect-square
condition \perfsq\ is satisfied). \HM\ then gets replaced with
\eqn\HMw{M_{J^{ia}\la,\tilde{J}^{jb}\tilde\mu}\ =\
\left\{\matrix{N_{\la}^{R_0\tilde\mu}\ &,\ \ \forall\ \la\in
P_+^0(A_{r,k}),\tilde\mu\in \tilde P_+^0(A_{k-1,r+1})\ ,\cr 0\ ,&\ \
{\rm otherwise\ \ ,}\cr}\right. } for all $i,j$.  The well-definedness
of $\cN$ in \HMw,
\HM\ and elsewhere is a consequence of \MJii, \MJiii.

For \HMw\ to be modular invariant, \chetA\ and \perfsq\ must be satisfied.
But anomalous invariants will be obtained whenever both
$k$ and $\r$ are odd, 3 divides $k\r$, and \perfsq\ is satisfied.  

Modular invariance can be verified easily by following the steps
used above for \HM. In this more general case, \SSMone\ has an extra
factor of $ab$. Then the right hand side of \SSMtwo\ has an extra factor
of $ab/a^2=b/a$, which exactly cancels the $\sqrt{\r/k}$. So
\krbar\ gets replaced with the perfect square condition \perfsq, and
we obtain infinitely many new heterotic invariants for each rank $r$.
That $\r$ and $k$ must both be odd, follows from $T$-invariance \Tinv\
applied to $M_{J^a0,\tilde{0}}=M_{0,\tilde{J}^b\tilde{0}}=1$.

The strategy used in constructing the heterotic invariant \HMw\
applies in a much more general setting: namely, one where the final 
holomorphic/antiholomorphic algebras are
\eqn\gensetf{\matrix{g_L=&su(\r_1)_{k_1}\+ su(k_1)_{\r_1}\+
\cdots\+ su(\r_m)_{k_m}\+ su(k_m)_{\r_m}\cr
g_R=&su(\s_1)_{\ell_1}\+ su(\ell_1)_{\s_1}\+\cdots\+
su(\s_n)_{\ell_n}\+ su(\ell_n)_{\s_n}}} where $\s_i=s_i+1$.
Condition \perfsq\ will be replaced here by more complicated constraints
most conveniently expressed in integral lattice terminology -- we won't
give the details here.

An example though is the construction of (nonheterotic) physical invariants
for $g_L=g_R=su(k)_\r$ from those of $h_L=h_R=su(\r)_k$. Choose any $a|\r$,
$ka^2/\r\in {\bf Z}$, and integers $b|a$ and $b'$ such that
$k(b^2-b'{}^2)/\r\in {\bf Z}$ (if $\r$ is even, replace these
denominators $\r$ with $2\r$). We require $\cN$ to be any physical invariant
with $(J^a;\id),(\id;J^a),(J^b;J^{b'})\in{\cal J}(N)$,
obeying the additional property that for any $\la,\mu\in
P_+(A_{r,k})$ with $N_{\la}^{\mu}\ne 0$, there exist simple currents
$(A;A')\in{\cal J}(N)$ such that $A\la,A'\mu\in P^0_+$. This will
happen provided $kab/\r$ is an integer coprime to $a$. Then we will be
able to define a nonheterotic invariant by dual-flipping: choose
$\tilde{a}|k$, $\tilde{b}|\tilde{a}$ and $\tilde{b}'$ satisfying
$\r\tilde{a}^2/k,\r(\tilde{ b}-\tilde{b}'{}^2)/k\in {\bf Z}$, and
$\r\tilde{a}\tilde{b}/k$ is an integer coprime to $\tilde{a}$. Then
\eqn\nonhet{
\tilde{N}_{\tilde{J}^{\tilde{a}\ell+\tilde{b}i}\la}^{\tilde{J}^{\tilde{a}j+
\tilde{b}'i}\mu}\ =\ N_{R_0\la}^{R_0\mu}} for any $\la\in P_+^0,\mu\in
P^0_+$, where all other entries of $\tilde{N}$ vanish. This always
works, unless both $k$ and $\r$ are even and the same power of 2
exactly divides each. If $\cN$ is a simple current invariant (i.e.  is
given by (5.6) below), then so is $\tilde{\cN}$. Much more interesting
is when $\cN$ is exceptional. For example, the level 3 exceptionals of
$A_r$ (given in \ref\tglmp{T. Gannon, {\sl Lett. Math. Phys.} {\bf 39}
(1997) 289}) are dual-shifts of the familiar exceptionals of $A_2$,
and the level 2 exceptionals (except the $A_{9,2}$ one, which succumbs
to the power of 2 condition) are dual-flips of the exceptionals of
$A_1$. Similar conditions apply to the $A_{3,8}\leftrightarrow
A_{7,4}$ and $A_{4,5}\leftrightarrow A_{4,5}$ exceptionals in
\tgades. It is possible to generalise this correspondence, and indeed
we describe at the end of section 5 a generalisation valid for any
$\r$, $k$ and $\cN$ (in particular, it allows us to recover the
$A_{9,2}$ exceptional from the $A_{1,10}$ one).  This correspondence
is clearly of value for the classification of the nonheterotic
partition functions of $A_{r,k}$.

\newsec{$C_r$, $B_r$ and $D_r$ invariants}

The other classical algebras can be treated in a straightforward
manner using the results of \mnrs.

$X_{r,k}=C_{r,k}$ is simplest. We have an order-two diagram symmetry
$J$ that acts on weights in the following way: \eqn\oaC{J\la\ =\
J(\sum_{i=1}^r\, \la_i w^i)\ =\ \sum_{i=1}^{r} \la_{r-i}w^i\ .} 
The $C_r$ version of \JSt\ is \eqn\JStC{S_{\la}^{J\mu}\ =\
\exp[\pi it_J(\la)]\, S_{\la}^{\mu}\ \ ,} with \eqn\tC{t_J(\la)\ =\
2(J^{-1}0)\cdot \lambda/k\ =\ 2w^r\cdot \lambda\ =\ \sum_{i=1}^r\,
i\la_i\ \ .} The action of $J$ on the elements of $T$ is given by
\eqn\JTC{T_{J\la}^{J\mu}\ =\ \exp\left[\pi i\left(
kr/2-t_J(\la)\right)\right]\, T_{\la}^{\mu}\ \ .}

Each weight in $P_+(C_{r,k})$ is in one-to-one correspondence with a
Young diagram with no more than $r$ rows, and no more than $k$
columns. For this algebra, the transpose operation $R$ supplies a
bijection between the weights of $P_+(C_{r,k})$ and
$P_+(C_{k,r})$. The modular matrices obey \eqn\SRC{\tilde
S_{R\la}^{R\mu}\ =\ S_{\la}^{\mu}\ \ ,\ \ \tilde T_{R\la}^{R\mu}\,
\tilde T_{\tilde 0}^{*\,\tilde 0}\ =\ T_{\la}^{*\,\mu}\, T_{0}^{0}\,
\exp\left[ \pi i t_J(\la) \right]\ \ .}
(The $C_{r,k}$ $S$-matrix is real, hence there is no `$*$' on $\tilde{S}$
in \SRC.)
Provided 4 divides $rk$, the analysis used for
$A_{r,k}$ goes through, using
 \eqn\HMC{M_{\tilde\mu,\nu}\ :=\
N_{\tilde\mu}^{R_0\nu}\ \ ,}
where $\cN$ is {\it any} nonheterotic invariant, and $R_0{\nu}:=
\tilde J^{{t}_J({\nu})}R{\nu}$. 
For a non-anomalous physical invariant, the central charge condition must
be satisfied. It becomes \eqn\ccC{c(C_{r,k}) + c(C_{k,r})\ =\ 2kr\
\equiv\ 0\ ({\rm mod}\ 24)\ \ ,} i.e. 12 must divide $kr$.

The generalisation corresponding to \gensetf\ is also possible
for the $C_r$ algebras. Furthermore, there is no need to stick to one
type of algebra. We can start with an invariant with algebras
\eqn\gengeni{h_L\,=\,h_R\,=\,
g_1\+g_2\+\ldots\+g_m\+h_1\+h_2\+\ldots\+ h_n\ } and find one with
\eqn\gengenf{\eqalign{ g_L\,=&\, g_1\+\tilde g_1\+ \ldots \+g_m\+\tilde
g_m\ \cr g_R\,=&\, h_1\+\tilde h_1\+ \ldots \+h_n\+\tilde h_n\ \ ,\cr }}
with some restrictions. But we will not belabour the point by making
this more explicit.

Level-rank duality for the orthogonal algebras, $B_r= so(2r+1)$ 
and $D_r= so(2r)$, is best treated together \mnrs. For example, if 
$so(N)_k$ denotes the affine $so(N)$ algebra at level $k$, the duality is 
\eqn\solr{so(N)_k\ \ \leftrightarrow\ \ so(k)_N\ \ ,}
and so does not respect the difference between the $B$ and $D$ algebras.  

When they exist, the duality equations for spinor representations of
the orthogonal algebras are much more complicated. For this reason, we
will restrict ourselves to taking duals of tensor representations
only.  The highest weights $\lambda$ of tensor representations of
$B_r$ satisfy \eqn\tensorB{\la_r\ \equiv\ 0\ ({\rm mod}\ 2)\ \ ,} and
those of $D_r$ obey \eqn\tensorD{\la_r\ \equiv\ \la_{r-1}\ ({\rm mod}\
2)\ \ .}  Let $P_+^0 (so(N)_k)$ denote the corresponding subset of
$P_+(so(N)_k)$. In both cases the condition can be written as
\eqn\tensorSO{t_J(\la)\ :=\ w^1\cdot\la\ \equiv\ 0\ ({\rm mod}\ 1)\ \
,} where $J$ is the following simple current:
\eqn\scBD{\eqalign{J\la&\,=\la_0w^1+\la_2w^2+\cdots+\la_rw^r\ \ \ {\rm
for} \ B_r\ , \cr
J\la&\,=\la_0w^1+\la_2w^2+\cdots+\la_{r-2}w^{r-2}+\la_rw^{r-1}
+\la_{r-1}w^r\ \ {\rm for}\ D_r\ .}}

There is one remaining complication for the case of $D_r$. The
irreducible tensor representations of $so(N)$ are in one-to-one
correspondence with Young diagrams having the sum of the lengths of
their first two columns less than or equal to $N$, except when the
first column is of length $N/2$. If $\la$ denotes a weight of $D_r$,
then let $C\la$ denote the weight obtained by exchanging the Dynkin
labels $\la_r$ and $\la_{r-1}$.  (For convenience define $C\la=\la$
for weights of $B_r$.)\foot{As defined here, $C$ is not the charge
conjugation for $D_r$ with $r$ even; it is for the other cases.} A
diagram with first column length $N/2=r$ corresponds to the pair
$[\la]:=\{\la,C\la\}$. Again the level-rank dual $R$ is given by the
transpose operation on Young diagrams. So it will be more convenient
to consider level-rank duality for the $C$-orbits $[\la]$. For this
purpose, define \eqn\directsumST{S_{[\la]}^{[\mu]}\ :=\
\sum_{{\kappa\in[\la]\atop\nu\in[\mu]}}S_{\kappa}^{\nu}\ ,\ \
T_{[\la]}^{[\mu]} \ :=\ \sum_{\kappa\in[\la]}T_{\kappa}^{\mu}\ .}

First, let us write the level-rank duality relation for $T$ \mnrs :
\eqn\lrTB{T_{[\la]}^{[\la]}\, T_{[0]}^{*\,[0]}\ =\ \exp\left[\pi
it'(\la)\right]\, \tilde T_{R[\la]}^{*\,R[\la]}\, \tilde T_{[\tilde
0]}^{[\tilde 0]}\ \ ,} valid for all $so(N)_k$, with
\eqn\tlaB{t'(\la)\ =\ \sum_{i=1}^{r-1}i\la_i + r\la_r/2\ \ , \ {\rm
for}\ so(N)_k= B_{r,k}\ \ ,} and \eqn\tlaD{t'(\la)\ =\
\sum_{i=1}^{r-2}i\la_i + {r-2\over 2}\la_{r-1}+{r\over 2}\la_r\ \ , \
{\rm for}\ so(N)_k= D_{r,k}\ \ .}  Notice that for a tensor
representation of highest weight $\la$, $t'(\la)\in \Z$. Also, for
$D_{r,k}$, $t'(\la)\equiv t_{J_s}(\la)$ or ${1\over 2}t_{J_s}(\la)$
(mod 2), for $r$ even or odd resp., where $J_s$ is the simple current
given by \eqn\scs{\eqalign{J_s\la=&\, \sum_{i=1}^r\la_{r-i}w^i\
\ \ {\rm for}\ r\ {\rm even}\ ,\cr J_s\la=&\,\la_rw^1
+\sum_{i=2}^r\la_{r-i}w^i\ \ {\rm for}\ r\ {\rm odd}\ .}}

The level-rank duality relation for $S$ is \mnrs :
\eqn\lrSSO{S_{[\la]}^{[\mu]}\ =\ \tilde S_{R[\la]}^{R[\mu]}\ ,\ \ \
\forall\, \la,\mu\in P_+^0(so(N)_k)\ \ .}

Incidently, the level-rank duality $so(N)_k\leftrightarrow so(k)_N$
extends in the natural way to small $N$, using $so(3)_k\cong
A_{1,2k}$, $so(4)_k\cong A_{1,k}\oplus A_{1,k}$, $so(5)_k\cong C_{2,k}$,
and $so(6)_k\cong A_{3,k}$.

Consider first $h_L=h_R=B_{r,k}$. We require the nonzero entries 
$N_\la^\mu$ of the initial invariant to obey $\la,\mu\in
P_+^0(B_{r,k})$, which means \eqn\condBD{N_{J0}^0=N_{0}^{J0}=1\ .}
This forces $k$ to be even, in which case such invariants will always
exist. When 16 divides $k$, we can get a
$B_{r,k}\oplus D_{k/2,2r+1}$ meromorphic invariant by putting
\eqn\merBD{M_{\la, J_s^jC^iR_0\mu}:=N_{\la}^\mu\ ,} where
$R_0\la:=J^{t'(\la)}R\la$, and $i,j\in\{0,1\}$ (the simple current
$J^{t'}$ cancels the sign picked up in \lrTB, but forces the presence
of the $J_s^j$, which is what requires the constraint
$16|k$). (Another way of cancelling the sign in \lrTB\ would be to use
instead $J_s^{t'}$, but again this requires $16|k$ and yields nothing
new: $M\tilde{C}$ instead of $M$.)

The other alternative is $h_L=h_R=D_{r,k}$ with $k$ even. As before, we
require the nonzero entries $N_\la^\mu$ to satisfy $\la,\mu\in P_+^0(D_{r,k})$.
Also, the fact that \lrSSO\ only holds at the level of $C$-orbits, means
that $C$ should commute with $N$. Thus we obtain the constraints
\eqn\condDD{N_{J0}^0=N_0^{J0}=1\ {\rm and}\ 
N_{\la}^\mu=N_{C\la}^{C\mu}
\qquad \forall \la,\mu\ .} Such invariants will always exist, 
since $k$ is even.

Now to cancel the sign in \lrTB, we must use $J_s^{t'}$, which
requires 8 to divide $rk$. The formula becomes
\eqn\merDD{M_{\la, J^jC^iR^0\mu}:=N_{\la}^\mu\ ,}
where $R^0\la:=J_s^{t'(\la)}R\la$ or $J_s^{2t'(\la)}R\la$, for $r$ even or odd
resp., and $i,j\in\{0,1\}$.

In both cases, the constraint on $N,k$ is that 16 divides $Nk$. For a
non-anomalous physical invariant,
 we need only impose the central charge condition
\eqn\ccSO{c(so(N)_k) + c(so(k)_N)\ =\ {Nk\over 2}\ \equiv\ 0\ ({\rm
mod}\ 24)\ \ ,} or 48 must divide $Nk$.

\newsec{Examples}

In order to write some simple examples of the new physical invariants
in a concise way, we introduce a little notation. Let
$[\la_0,\la_1,\ldots,\la_r]$ represent the {\it character} ${\rm
ch_\la}$, with $\la=\sum_{i=1}^r\la_iw^i$, where we put
$\la_0:=k-\sum_{i=1}^r\la_ia^\vee_i$ as above. So, the trivial
diagonal physical invariant for algebra $X_{r,k}$ would be written as
\eqn\diag{\sum_{\la\in P_+(X_{r,k})}\,
\lv[\la_0,\la_1,\ldots,\la_r]\rv^2\ } in this notation.

We will content ourselves here with simple examples involving the
$A_r$ and $C_r$ algebras only.

A large class of nonheterotic invariants for $X_{r,k}$, with
$X_r$ simple, is the class of partition functions of orbifolds by
subgroups of the centre of $\exp(X_r)$ \ber\ref\alzfgk{D. Altschuler,
J. Lacki and P. Zaugg, {\sl Phys. Lett.} {\bf B205} (1988)
281\hfill\break G. Felder, K. Gawedzki and A. Kupiainen, {\sl
Comm. Math. Phys.} {\bf 117} (1988) 127}. Let us first discuss
meromorphic invariants that can be
constructed from them.

A manifestly non-negative formula was written in \ref\ahnw{C. Ahn and
M.A. Walton, {\sl Phys. Lett.} {\bf B223} (1989) 343} for these 
invariants: \eqn\ZNall{N(\exp(X_r)/Z_n)_{\mu}^{\lambda}\ =\
\sum_{m=0}^{n-1}\, \delta_{\mu}^{A^m\lambda} \ \delta_1\{\,
(Aw^0)\cdot\left(\lambda+{1\over 2}mkAw^0\right) \,\}\ \ .}  It is
valid for any cyclic orbifold group $Z_n$, as long as
\eqn\condZ{{{nk(Aw^0)\cdot (Aw^0)}\over 2}\ \equiv\ 0\, ({\rm mod}\
1)\ \ .}  Here $A$ generates the corresponding diagram symmetry
subgroup, so that $|A|=n$, and \eqn\dxy{\delta_y\{x\}\ =\
\left\{\matrix{1\ \ ,&{\rm if}\ x/y\ {\rm is\ integer;}\ \cr 0\ \
,&{\rm otherwise.}\cr} \right.\ }

First consider nonheterotic $A_{r,k}$ physical invariants
$\cN=\sum_{\la,\mu}\, {\rm ch}_\la\, N_{\la}^{\mu\,} {\rm ch}_\mu^*$
of this orbifold type. For $A_{r,k}$, the outer automorphism in
\ZNall\ can be $A=J^d$ for any $d$ dividing $\r$. However, in order to
construct the simplest meromorphic physical invariants $\cM
=\sum_{\la,\mu}\, {\rm ch}_\la\, {\rm ch}_\mu\, M_{\la,\mu}$, i.e.\
using \HM, we have seen in \Mcond,\krbar\ that we must put $A=J$, and
also take $k=\r$ odd. (We will consider the more general \HMw\ later.)

First, we will ignore the central charge condition \cheta, and find a
simple anomalous meromorphic physical invariant. The lowest value of $k=\r$
at which a non-diagonal invariant \ZNall\ exists is 3. Eqn. \ZNall\
yields \eqn\zAone{ \cN\ =\ \left| [3,0,0]+[0,3,0]+[0,0,3] \right|^2\,
+\, 3\left| [1,1,1]\right|^2\ \ ,} and \HM\ produces \eqn\smallHaA{
\cM^{(an.)}\ =\ \left( [3,0,0]+[0,3,0]+[0,0,3] \right)^2\, +\, 3\left(
[1,1,1]\right)^2\ \ .}

For a non-anomalous physical invariant, \cheta\ requires $k^2\equiv
1({\rm mod}\ 24)$. But this just means that $k=\r$ must be coprime to
6, so that the construction works for all even $r$ that are not
congruent to 2 (mod 3).  An {\it infinite series} of meromorphic
physical invariants is thus obtained. The smallest possible rank is
$r=4$, where \eqn\zAtwo{
\eqalign{\cN\ =\ |J\{5,0,0,0,0\}|^2 +
|J\{3,1,0,0,1\}|^2 +& |J\{3,0,1,1,0\}|^2 \cr
+ |J\{1,2,0,0,2\}|^2 + |J\{1,0,2,2,0\}|^2 +&
5|[1,1,1,1,1]|^2\ ,}} where we have defined 
\eqn\orb{A\{\la_0,\la_1,\ldots,\la_r\}\
:=\ \sum_{a=1}^{|A|}\,
A^a\,[\mu_0,\mu_1,\ldots,\mu_r]\ .} 

In general, $R$ is a bijection between the set of $J$-orbits in
$P_+(A_{r,k})$ and the set of $\tilde J$-orbits in $P_+(A_{k-1,r+1})$.
$R_0$ is a bijection between $P_+^0(A_{r,k})$ and $\tilde
P_+^0(A_{k-1,r+1})$. If $D:=gcd\{\r,k\}$, $P_+^0(A_{r,k})$ and $\tilde
P_+^0(A_{k-1,r+1})$ are composed of complete $J^{\r/D}$-orbits and
$\tilde J^{k/D}$-orbits, respectively. So, when $\r=k$, $R_0$ maps the
$J$-orbits of $P_+^0(A_{r,k})$ onto the $\tilde J$-orbits of $\tilde
P_+^0(A_{k-1,r+1})$. Consequently, the meromorphic $(A_{4,5}\oplus
A_{4,5};\{\})$ invariant that results from \zAtwo\ is \eqn\HAtwo{
\eqalign{\cM\ =\ J\{5,0,0,0,0\}^2 +
J\{3,1,0,0,1\}^2 +&\,J\{3,0,1,1,0\}\,J\{1,2,0,0,2\} \cr
+J\{3,0,1,1,0\}\, J\{1,2,0,0,2\} +& J\{1,0,2,2,0\}^2 +\,
5[1,1,1,1,1]^2\ ,}} which corresponds to the 67th entry in the
table of $c=24$ meromorphic theories given in \schell.

The invariants for larger $\r=k$ are too long to write here, so we
will instead write a general, formal equation. Let
$P_{+,J}(A_{r,\r})$ denote a set of representatives of the $J$-orbits
in $P_+^0(A_{r,\r})$. Then the general form of the nonheterotic
$A_{r,\r}$ physical invariant is \eqn\genMA{\cN\ =\ \sum_{\la\in
P_{+,J}(A_{r,\r})}\ {\r\over{\Vert\langle\la\rangle_J\Vert}} \ \left|
\, \langle {\rm ch}_\la\rangle_J\,\right|^2\ \ ,} with $\langle
\la\rangle_J$ being the $J$-orbit of $\la$. By \HM, we construct the 
following meromorphic physical invariant \eqn\genHA{\cM\ =\
 \sum_{\la\in P_{+,J}(A_{r,\r})}\
 {\r\over{\Vert\langle\la\rangle_J\Vert}} \ \left( \, \langle {\rm
 ch}_\la\rangle_J\,\right)\, \lb\langle \widetilde{\rm
 ch}_{R_0\la}\rangle_{
\tilde{J}} \,\rb \ ,} as long as \cheta\ is satisfied.

An example of an (anomalous) invariant of the type \HMw\ is also too
long to write here, but a general equation similar to \genMA\ can
easily be found.

So far we have only considered examples of meromorphic invariants that
can be obtained from the nonheterotic orbifold invariants
\ZNall. Thus we have not yet considered any example of an exceptional
invariant. There is no obstruction to using such invariants,
however. Consider the exceptional invariant of
$(A_{4,5};A_{4,5})$ \tgades: \eqn\excM{\eqalign{\cN\ =\
&\lv J\{0,0,5,0,0\}\rv^2 + J\{0,1,3,1,0\}\,[1^5]^* +
[1^5]\,J\{0,1,3,1,0\}^* +\cr &\lv J\{1,0,3,0,1\}\rv^2 +
\lv J\{0,2,1,2,0\}\rv^2 + \lv J\{2,0,1,0,2\}\rv^2 +
4\lv[1^5]\rv^2\ \ .}}  The invariant obtained by \HM\ is then just
\eqn\excH{\eqalign{\cM\ =\ &\left(J\{0,0,5,0,0\}\right)^2 +
\,J\{0,1,3,1,0\}\,[1^5] + [1^4]\,J\{0,1,3,1,0\}+\cr
&\,J\{1,0,3,0,1\}\,J\{0,2,1,2,0\} +J\{0,2,1,2,0\}\,
J\{1,0,3,0,1\}+\cr&\,
\left(J\{2,0,1,0,2\}\right)^2 + 4\left([1^5]\right)^2\ \ .}}
This is the ninth entry in the table of $c=24$ meromorphic theories 
given in \schell.

Next we treat $C_{r,k}$, where 4 divides $rk$. Any physical invariant
$\cN$ will generate a (possibly anomalous) meromorphic
invariant. Generically, there will be two physical invariants ${\cal
N}$ for such $r,k$: they correspond to the choices $A=id$ and $A=J$
in \ZNall. The resulting $\cM$ will be anomalous unless the central
charge condition \ccC\ is satisfied, i.e. unless 12 divides $kr$.

Four simple anomalous heterotic invariants can be found for
$kr=4$.  With $r=k=2$, \ZNall\ gives the
$(C_{2,2};C_{2,2})$ invariant \eqn\MCone{\cN\ =\ \left|
[2,0,0]+[0,0,2]\right|^2+2\left|[0,2,0]\right|^2+2\left|[1,0,1]\right|^2\
.} From this, the anomalous 
$(C_{2,2}\oplus C_{2,2};\{\})$ invariant 
\eqn\HaCone{\cM^{(an.)}\ =\ \left( [2,0,0]+[0,0,2]\right)^2+2[0,2,0][1,0,1]
+2[1,0,1][0,2,0]\ }
is obtained. For $\cN$ the diagonal invariant of $C_{2,2}$, we get
\eqn\HaCthr{\eqalign{\cM^{(an.)}\ =&\ [2,0,0]^2+[0,0,2]^2+[1,1,0][0,1,1]+
[0,1,1][1,1,0]\cr &+[1,0,1][0,2,0]+[0,2,0][1,0,1].\ }}
With $r=4,k=1$, we find
\eqn\MCtwo{\cN\ =\ \left|[1,0,0,0,0]+[0,0,0,0,1]\right|^2 +\ 
2\left|[0,0,1,0,0]\right|^2\ ,}
and the resulting anomalous $(C_{4,1}\oplus A_{1,4};\{\})$ invariant
\eqn\HaCtwo{\cM^{(an.)}\ =\ \left([1,0,0,0,0]+[0,0,0,0,1]\right)
\left([4,0]+[0,4]\right)\ +2[0,0,1,0,0][2,2]\ \ .}
Choosing $\cN$ diagonal instead gives us
\eqn\HaCfou{
\cM^{(an.)}\ =\ [1,0^4][4,0]+[0,1,0^3][1,3]+[0,0,1,0,0][2,2]+
[0^3,1,0][3,1]+[0^4,1][0,4]\ .}

The lowest value of $kr$ at which non-anomalous physical invariants can be
written is 12. As an example, we choose $r=4,k=3$. Eqn. \ZNall\ with
$A=J$ yields
\eqn\MCthree{\eqalign{\cN\ =\ \lv J\{3,0,0,0,0\}\rv^2\ +&\
2\lv[0,0,3,0,0]\rv^2 \cr +\ \lv J\{2,0,1,0,0\}\rv^2\ +&\
\lv J\{2,0,0,0,1\}\rv^2 \cr +\
\lv J\{1,2,0,0,0\}\rv^2\ +&\
\lv J\{0,2,1,0,0\}\rv^2 \cr +\
\lv J\{1,0,0,2,0\}\rv^2\ +&\
\lv J\{1,0,2,0,0\}\rv^2 \cr +\
\lv J\{1,1,0,1,0\}\rv^2\ +&\ 2\lv[0,1,1,1,0]\rv^2 \cr +\
2&\lv[1,0,1,0,1]\rv^2\ \ .  }} The meromorphic $C_{4,3}\oplus C_{3,4}$
physical invariant we
obtain from it is \eqn\HCone{\eqalign{\cM\ =\
J\{3,0,0,0,0\}\,\tilde J\{4,0,0,0\}\ +&\
2[0,0,3,0,0]\, [2,0,0,2]\cr +\
J\{2,0,1,0,0\}\,\tilde J\{2,2,0,0\}\ +&\
J\{2,0,0,0,1\}\,\tilde J\{0,4,0,0\}\cr +\
J\{1,2,0,0,0\}\,\tilde J\{3,0,1,0\}\ +&\
J\{0,2,1,0,0\}\,\tilde J\{2,1,0,1\}\cr +\
J\{1,0,0,2,0\}\,\tilde J\{1,0,3,0\}\ +&\
J\{1,0,2,0,0\}\,\tilde J\{2,0,2,0\}\cr +\
J\{1,1,0,1,0\}\,\tilde J\{1,2,1,0\}\ +&\
2[0,1,1,1,0]\, [1,1,1,1]\cr +\
2[1,0,1,0,1]\,&[0,2,2,0]\ \ ,  }}
which corresponds to the 70th entry in \schell. 

Notice that the algebra $C_{4,3}\oplus C_{3,4}$ does not satisfy a
condition of \schell: that the ratio of the dual Coxeter number to the
level be equal for all $X_{r,k}$ in the affine algebra. But that is
because the condition is derived assuming that the affine algebra is
maximal. Clearly, it is satisfied by the $D_{24,1}$ algebra in which
$C_{4,3}\oplus C_{3,4}$ is conformally embedded. We will explain how
\HCone\ can be recovered from this embedding at the end of the next
section.

\newsec{Level-one holomorphic physical invariants and progeny}
  
There is an intimate connection between level-rank dualities and
certain conformal embeddings. For example, the conformal subalgebra
$su(\r)_k\oplus su(k)_\r\subset su(\r k)_1$ contains much information
about the duality $su(\r)_k\leftrightarrow su(k)_\r$. So it is natural
to suspect that some of the invariants we find by dual-flipping may
also be obtainable (less directly) by the method of conformal
subalgebras.

But the simplest of our invariants are meromorphic. To derive them
using conformal subalgebras, we would need to start with meromorphic
level-one physical invariants for $X_r$, i.e. physical invariants for
$(X_{r,1};\{\})$. Previous to this work, only a few such invariants
were known -- for some of these, see e.g.\ \schell.

Motivated by this situation, however, we were led to derive such
invariants. Furthermore, we were able to classify all {\it heterotic}
invariants for algebras $(X_{r,1};\{\})$, $(X_{r,1};X_{s,1})$ and
$(X_{r,1}\oplus X_{s,1};\{\})$. The complete list for $(X_{r,1};\{\})$
follows, using the notation ${\rm ch}_i:={\rm ch}_{w^i}$.

\noindent(a)\ $\underline{(A_{r,1};\{\}),\ \r=s^2, \r\ {\rm odd}}:$\qquad
(anomalous unless 24 divides $r$)\hfill
$$\cM\ =\ {\rm ch}_0+{\rm ch}_s+{\rm ch}_{2s}+\ldots +{\rm ch}_{(s-1)s}\ .$$

\noindent(d)\ $\underline{(D_{r,1};\{\}),\ r\equiv 0\ (\mod 8)}$:\qquad
(anomalous unless 24 divides $r$)
\hfill $$\cM_1\ =\ {\rm ch}_0+{\rm ch}_r\ ,$$ 
$$\cM_2\ =\ {\rm ch}_0+{\rm ch}_{r-1}\ .$$ \noindent ${\rm ch}_r,{\rm
ch}_{r-1}$ are the highest weights of the spinor and conjugate-spinor
representations of $D_{r,1}$, respectively.

\noindent(e8)\ $\underline{(E_{8,1};\{\})}$:\qquad 
$\cM={\rm ch}_0$ is anomalous.

\noindent($*$)\ $\underline{(X_{r,1};\{\})\  
{\rm for}\ X_r=B_{r},C_r,E_6,E_7, F_4,G_2}$: \qquad no invariants.

The completeness of the above list follows immediately from
the simple level-one nonheterotic (i.e. $(X_{r,1};X_{r,1})$)
classification \ref\itzg{C. Itzykson, {\sl Nucl. Phys.}
Proc. Suppl. {\bf 5B} (1988) 150\hfill\break\noindent T. Gannon,
{\sl Nucl. Phys.} {\bf B396} (1993) 708}: if $\sum {\rm ch}_i$ is a
heterotic invariant (possibly anomalous), then $|\sum {\rm ch}_i|^2$ is a
nonheterotic physical invariant. So we can simply run down the
list of \itzg, to see if any of those invariants can be written as a
perfect square.

We will quickly run through the classification for $(X_{r,1}\+ X_{s,1};\{\})$
and $(X_{r,1};X_{s,1})$. Its proof is also straightforward; $T$-invariance
takes one almost all the way.

\noindent(a1)\ $\underline{(A_{r,1};A_{s,1})}$ has invariants provided
$\sqrt{\r\s}\in{\bf Z}$ and $r\equiv s$ (mod 2); they will be anomalous
unless $r\equiv s$ (mod 24).

\noindent(a2)\ $\underline{(A_{r,1}\+A_{s,1};\{\})}$ has invariants provided
$\sqrt{\r\s}\in{\bf Z}$, both $\r$ and $\s$ are odd, and every prime $p\equiv
3$ (mod 4) divides $\r$ exactly an even number of times; they will be
anomalous unless $r\equiv -s$ (mod 24).

\noindent(b1)\ $\underline{(B_{r,1}\+B_{s,1};\{\})}$ has exactly 1 invariant,
 provided $r+s\equiv 7$ (mod 8); it will be anomalous unless $r+s\equiv -1$
 (mod 24).

\noindent(b2)\ $\underline{(B_{r,1};B_{s,1})}$ has exactly 1
invariant, provided $r\equiv s$ (mod 8); it will be anomalous unless
$r\equiv s$ (mod 24).
 
\noindent(d1)\ $\underline{(D_{r,1}\+D_{s,1};\{\})}$ will have no invariants,
unless $r\equiv -s$ (mod 8); it will have exactly 6 if either $r\equiv s\equiv
0$ or $r\equiv s\equiv 4$ (mod 8), otherwise it will have exactly 2 provided
$r\equiv -s$ (mod 8)); all of these will be anomalous unless $r\equiv -s$ (mod
24).

\noindent(d2)\ $\underline{(D_{r,1};D_{s,1})}$ will have no invariants,
unless $r\equiv s$ (mod 8); it will have exactly 6 if either $r\equiv s\equiv
0$ or $r\equiv s\equiv 4$ (mod 8), otherwise it will have exactly 2 (provided
$r\equiv s$ (mod 8)); all of these will be anomalous unless $r\equiv s$ (mod
24).

The only other heterotic invariant\foot{\small For the choice $X=C$,
we have been able to show this only by assuming the invariant is
non-anomalous, i.e.\ obeys \clcr, although we expect there will be no
anomalous heterotics either.}  of type $(X_{r,1}\+ X_{s,1};\{\})$ or
$(X_{r,1};X_{s,1})$ for $X$ simple, is the anomalous one $\cM={\rm
ch}_0{\rm ch}_0$ for $X_r= X_s=E_8$.

The invariants in (b1), (b2), (d1) and (d2) are easy to write down. Those in
(a1) and (a2) are much messier and should be interpreted in the language 
of self-dual lattices.

Some of the invariants found above seem to be partition functions for
strings on heterotic nonsimply-connected group manifolds. That is, as
partition functions they appear to describe closed bosonic strings
with left-moving sector and right-moving sector propagating on {\it
different} nonsimply-connected group manifolds. Perhaps the simplest
example is (a), where the group would be $SU(s^2)/Z_s$, and the string
meromorphic.

We want to show that invariants found in the previous section by
dual-flipping can also be found using conformal subalgebras and
contraction. The appropriate conformal subalgebras are what we dub the
{\it level-rank} conformal subalgebras, for obvious reasons: the
prototype is $su(m)_n\oplus su(n)_m\subset su(mn)_1$. In this work, we
will consider only this level-rank conformal subalgebra.

First let us write down its conformal branching rules \mw\abi. Let us
again use the notation of the last section, where a weight denotes its
corresponding affine character. Then we have
\eqn\confbr{w^j\ =\ \sum_{{\la\in
P_+(A_{m-1,n})\,:\,}\atop{t_J(\la)\equiv j\,(\mod m)}}\ \la\,(\tilde
J^{\,\left[j-t_J(\la)\right]/m}\, R\la)\ \ ,} where here $w^j$ denotes
the character of the $su(mn)_1$ representation restricted to the
subalgebra. This last formula is consistent under the swap of
$su(m)_n$ and $su(n)_m$, as must be. That's because $\tilde
t_J(R\la)\equiv t_J(\la)\ (\mod n)$ and \tJA\ imply \eqn\ttj{\tilde
t_J(\tilde J^{[j-t_J(\la)]/m}\, R\la )\ \equiv\ j\, (\mod n)\ \ .}

With this result we will be able to show that the level-one invariant
of (a) above yields the meromorphic invariants \HM\ and \HMw\ for
$\cN$ given by \ZNall\ with $A=J$ and $A=J^a$,
respectively. Substituting \confbr, with $m=n=s$, into the
$(su(s^2)_1;\{\})$ invariant $\sum_{a=0}^{s-1}\, w^{as}$ yields
\eqn\HHi{\sum_{a=0}^{s-1}\ \sum_{\la\in P_+^0(A_{s-1,s})}\ \la\,\left(
J^{a-t_J(\la)/s} \, R\la \right)\ =\ \sum_{\la\in P_+^0(A_{s-1,s})}\
\la\ \left( \sum_{a=0}^{s-1}\ J^a\, R_0\la \right)\ ,} using
\Rzero. But this equals \genHA\ with $\r=s$. More generally, when $\r
k=s^2$, this construction yields \HMw\ for $\cN$ given by the choice
$A=J^a$ in \ZNall.

To recover other invariants, such as \excH, we must also use the
technique of modular invariant contraction \bou: that
is, contracting one modular invariant with a second produces another
modular invariant. For example, let $M'_{\la,\nu}$ be
 the coefficient matrix of the meromorphic $(A_{4,5}\oplus
A_{4,5};\{\})$ invariant \HAtwo, a special case of \HHi, just obtained
by conformal embedding. Then, if $N_{\la}^{\nu}$ is 
the matrix for the exceptional $(A_{4,5};A_{4,5})$ invariant \excM,
the contraction \eqn\cont{{1\over 5}\sum_{\nu\in P_+(A_{4,5})}\
M'_{\la,\nu}\, N_{\mu}^{\nu}\ =\ M_{\la,\mu}} yields the matrix
elements of the exceptional $(A_{4,5}\oplus A_{4,5};\{\})$ invariant
\excH. The normalisation ${1\over 5}$ in \cont\ is needed to get
$M_{0,0}=1$; in general this necessary step could introduce fractions
into the modular invariant. The possibility of fractions is the main
difficulty with any contraction formula.

It appears that any invariant obtainable by the dual-flip can also be
recovered from a level-one invariant by the combination of level-rank
conformal subalgebras and modular invariant contraction. We have found
no exception to this rule. For example, consider the large set of
$(su(k)_\r;su(k)_\r)$ invariants described as the dual-flips of
$(su(\r)_k;su(\r)_k)$ invariants by \nonhet. We will follow the
procedure given in \mw.  Start with the diagonal
$(su(k\r)_1;su(k\r)_1)$ invariant, and apply the conformal embedding
$su(\r)_k\oplus su(k)_\r \subset su(k\r)_1$, to get an invariant $\cN''$
for $(su(\r)_k\+ su(k)_{\r};su(\r)_k\+ su(k)_{\r})$. Let $\cN$ be any
$(su(\r)_k;su(\r)_k)$ physical invariant, and let $\ell$ be the
smallest number $1\le \ell\le\r$ for which $N_{J^\ell 0}^{J^\ell 0}\ne
0$. Define the contraction
\eqn\contr{\widetilde{N}_{\mu}^{\mu'}={\ell\over \r}\sum_{\la,\la' \in
P_+(A_{r,k})}N_{\la,\mu}^{\prime\prime\,\la',\mu'}\,N_{\la'}^{\la}\ .}
It turns out (using \MJii,\MJiii) that with the normalisation
${\ell\over \r}$, $\widetilde{\cN}$ will be a {\it physical invariant}
for $(su(k)_{\r};su(k)_{\r})$ (this could only be conjectured in \mw).

However, {\it there is no bijection between the $(su(k)_{\r};
su(k)_{\r})$ and $(su(\r)_k;su(\r)_k)$ physical invariants.} This is
already evident in \tglmp. What we obtain here is just as
important. It is one of the main results of the paper, and so we will
state it more precisely.  We call a physical invariant $\cN$ a {\it
simple current invariant} if it obeys the selection rule
\eqn\sci{N_{\la}^{\mu}\ne 0\ \Rightarrow\ \la=A\mu\ {\rm for\ some\
simple\ current}\ A} ($A$ depends on $\la,\mu$). We will call a
physical invariant $\cN$ {\it exceptional} if neither $\cN$ nor any
conjugation $C_1\cN C_2$ is a simple current invariant (for $A_r$, the
only nontrivial conjugation is the charge conjugation $C$, defined by
$(C\la)_i=\la_{\r-i}$).

\medskip \noindent{\it Fact}\quad (a) If $\cN$ is a physical invariant
for $(su(\r)_k;su(\r)_k)$, then $\widetilde{\cN}$ defined by \contr\
will be a physical invariant for $(su(k)_{\r};su(k)_{\r})$.

\smallskip\noindent{(b)} Suppose $su(\r)_k$ has an exceptional invariant
$\cN$ obeying $N_{J0}^{J0}\ne 0$. Then $su(k)_{\r}$ will have an exceptional
invariant obeying $\widetilde{N}_{\tilde{J}\tilde{0}}^{\tilde{J}\tilde{0}}\ne
0$.\medskip

More generally, if $N_{\la}^{\la'}\ne 0$, $t_J(\la)\equiv t_J(\la')$
(mod $\r$), and $\la'\ne J^i\la$ for any $i$ (respectively, $\la'\ne
C^iJ^j\la$ for any $i,j$), then $\widetilde{N}$ will have the
corresponding property.  This should be of considerable value in
future attempts at $A_{r,k}$ physical invariant classifications.  Such
weights $\la,\la'$ can be found for any known exceptional invariant of
$A_{r,k}$. So we can confidently predict the following (which we know
to be true if either $\r$ or $k$ is a product of distinct odd primes):

\medskip\noindent{\it Conjecture}\quad $(su(\r)_k;su(\r)_k)$ will have an
exceptional physical invariant, iff $(su(k)_{\r};su(k)_{\r})$ does. \medskip

On the other hand, there {\it is} a bijection between the nonheterotic
physical invariants of $C_{r,k}$, and those of $C_{k,r}$, given by
the formula $N_\la^\mu=\widetilde{N}_{R\la}^{R\mu}$. 

As promised, we conclude this section with a comment on the c=24
invariant \HCone. As stated previously, it corresponds in Schellekens'
table \schell\ to the 70th entry, with algebra $D_{24,1}$. The two
invariants of (d) above, with r=24, differ by a conjugation (not
charge conjugation) and also correspond to entry 70. The invariant
\HCone\ can be obtained from one of them by the conformal embedding
$C_{4,3}\oplus C_{3,4}\subset D_{24,1}$, with branching rules given in
\dv. The other (d) invariant produces a different $(C_{4,3}\oplus
C_{3,4};\{\})$ invariant. But it too can be derived by the dual-flip,
this time from the diagonal non-heterotic invariant of either
$C_{4,3}$ or $C_{3,4}$.

\newsec{Conclusion}

To summarise, we have shown how many new heterotic physical invariant
combinations of affine Kac-Moody characters can be found. The new
invariants are intimately related with the level-rank duality of
affine Kac-Moody algebras. They can be derived either of two ways: (i)
by the new dual-flip method (see e.g. \HM), or (ii) by applying
together the two old methods of conformal embeddings \bt\ and modular
invariant contraction \bou\ (by nonheterotic invariants).

The advantage of the first method is that it is the most direct and explicit.
To use the old method (ii), it was first necessary to find new level-one
heterotic invariants by force. Method (i) also shows explicitly that
level-rank duality is responsible for the existence of the new
invariants. This is satisfying, since other generic symmetries of the
modular matrices, such as Galois \fss\ and simple-current relations
\sy\int, have been shown to produce invariants.

Level-rank duality appears in method (ii) when those conformal
subalgebras related to it (such as $A_{m-1,n}\oplus A_{n-1,m}\subset
A_{mn-1,1}$) are used. But this points out an advantage of the second
method: conformal subalgebras besides the `level-rank' ones can be
used. In this sense, method (ii) is more general. We have not
investigated this generalisation here, except for the following
remark. The heterotic invariants in (a) and (d) above, together with
conformal embeddings such as \ref\confemb{F.A.\ Bais and P.G.\
Bouwknegt, {\sl Nucl.\ Phys.}\ {\bf B279} (1987) 561;
\hfill\break\noindent A.N.\ Schellekens and N.\ Warner, {\sl Phys.\
Rev.}\ {\bf D34} (1986) 3092} $A_{n,n-1}\subset A_{(n-1)(n+2)/2,1}$,
$B_{n,2}\subset A_{2n,1}$, $A_{2n,2n+1}\subset D_{2n(n+1),1}$,
$B_{2n,4n-1}\subset D_{n(4n+1)}$, $C_{2n,2n+1}\subset D_{n(4n+1),1}$,
$D_{2n,4n-2}\subset D_{n(4n-1),1}$, give us new meromorphic invariants
of type $(X_{r,k};\{\})$ (e.g.\ for $A_{49,48}$ and $A_{16,17}$).

We should emphasise, however, that compared to nonheterotic
invariants, heterotic invariants are rare, as the results of
\tghet\tgqhk\ show. In contrast, the dual-flip seems to produce a
large class of heterotic invariants. So, although we made no attempt
at completeness in this paper, it may still be that this method can
produce most heterotic invariants for the algebras with level-rank
duality. A natural question then is, have we exhausted all generic
ways of constructing heterotics?

This question is not premature, considering what seems to be the main
(and unexpected) lesson being learned from the nonheterotic physical
invariant classifications: the `obvious' ways to construct these
physical invariants (most notably, conjugations, simple currents,
conformal embeddings, and rank-level duality) succeed in constructing
almost all of them.

A direct challenge to this optimism is provided by the meromorphic CFTs,
when 8 divides $c$. There is at least one of these associated with each
even self-dual lattice of dimension $c$. In dimension 24, there are only
24 such lattices, but in dimension 32 there are at least 8 million, and
this lower bound grows quickly with the dimension. So there will
be an enormous number of $c=48$ meromorphic CFTs, very few of which can be
constructed using these ``generic'' methods. However, these meromorphic
CFTs will in general {\it not} be of the WZW type considered in this paper.
Our point is merely that if one restricts to an affine algebra, and varies
the level, then what one finds is very few if any truly exceptional
physical invariants. The intractability of an explicit classification
of all meromorphic CFTs, which is inherited from the intractability of an
explicit classification of all even self-dual lattices, has led many
(prematurely, in our opinion) to regard the WZW physical invariant
classification as hopeless.

\vskip 1truecm \noindent{\it Acknowledgements}\hfill\break MW thanks
Matthias Gaberdiel for helpful conversations and the High Energy
Physics group of DAMTP for hospitality.

\listrefs
\bye